\documentclass[11pt]{article}
\usepackage{amsfonts}
\usepackage{graphicx}
\usepackage{amsmath}
\usepackage{amssymb}
\usepackage{amsxtra}

\input{tcilatex}

\begin{document}

\title{Discreteness of point charge in nonlinear electrodynamics}
\author{A.I. Breev \\
breev@mail.tsu.ru, Tomsk State University, 36 Lenin\\
Prospekt, 634050, Tomsk, Russia\\
Tomsk Polytechnic University, Lenin avenue 30, Tomsk 634050, Russia \and %
A.E.Shabad \\
shabad@lpi.ru, P. N. Lebedev Physical Institute, 53 Leninskiy prospekt,\\
119991, Moscow, Russia\\
Tomsk State University, 36 Lenin Prospekt, 634050, Tomsk, Russia}
\maketitle

\begin{abstract}
We consider two point charges in electrostatic interaction between them
within the framework of a nonlinear model, associated with QED, that
provides finiteness of their field energy. We argue that if the two charges
are equal to each other the repulsion force between them disappears when
they are infinitely close to each other, but remains as usual infinite if
their values are different. This implies that within any system to which
such a model may be applicable the point charge is fractional, it may only
be $2^n$-fold of a certain fundamental charge, $n = 0,1,2...$

We find the common field of the two charges in a dipole approximation, where
the separation between them is much smaller than the observation distance.
\end{abstract}

\section{Introduction}

\label{s:intro} Introduction

Recently a class of nonlinear electrodynamic models was proposed \cite%
{CosGitSha2013a} wherein the electrostatic field of a point charge is, as
usual, infinite in the point where the charge is located, but this
singularity is weaker than that of the Coulomb field, so that the space
integral for the energy stored in the field converges. In contrast to the
Born-Infeld model, the models from the class of Ref. \cite{CosGitSha2013a}
refer to nonsingular Lagrangians that follow from the Euler-Heisenberg (E-H)
effective Lagrangian \cite{Heisenberg} of QED truncated at any finite power
of its Taylor expansion in the field. This allows us to identify the
self-coupling constant of the electromagnetic field with a definite
combination of the electron mass and charge and to propose that such models
may be used to extend QED to the extreme distances smaller than those for
which it may be thought of as a perfectly adequate theory. More general
models based on the Euler-Heisenberg Lagrangian, but fit also for
considering non-static nonlinear electromagnetic phenomena, where
not-too-fast-varying in space and time fields are involved, received
attention as well. Among the nonlinear effects studied, there are the linear
and quadratic electric and magnetic responses of the vacuum with a strong
constant field in it to an applied electric field \cite{AdoGitSha2016} ,
with the emphasis on the magneto-electric effect \cite%
{GitSha2012,AdoGitSha2013, AdoGitSha2014} and magnetic monopole formation 
\cite{AdoGitSha2015}. Also self-interaction of electric and magnetic dipoles
was considered with the indication that the electric and magnetic moments of
elementary particles are subjected to a certain electromagnetic
renormalization \cite{CosGitSha2013} after being calculated following a
strong interaction theory, say, QCD or lattice simulations. Interaction of
two laser beams against the background of a slow electromagnetic wave was
studied along these lines, too \cite{King}.

In the present paper we are considering the electrostatic problem of two
point charges that interact following nonlinear Maxwell equations stemming
from the Lagrangian of the above \cite{CosGitSha2013a} type, their common
field not being, of course, just a linear combination of the individual
fields. The problem is outlined in the next Section \ref{NME}. Once the
field energy is finite we are able to define the attraction or repulsion
force between charges as the derivative of the field energy with respect to
the distance $\mathbf{R}$ between them. Contrary to the standard linear
electrodynamics, this is evidently not the same as the product of one charge
by the field strength produced by the other! Based on the permutational
symmetry of the problem that takes place in the special case where the
values of the two charges are exactly the same we establish that the
repulsion force between equal charges disappears when the distance between
them is zero. This may shed light to the ever-lasting puzzle of whether a
point-like electric charge may exist without flying to pieces due to mutual
repulsion of its charged constituents. The optional answer proposed by the
present consideration might be that after admitting that these exists a
certain fundamental charge $q,$ every other point charge should be
fractional, equal to $2^{n}q,$ with $n,$ being zero or positive integer. In
Section \ref{Appr} we are developing the procedure of finding the solution
to the above static two-body problem in the leading approximation with
respect to the ratio of the distance $R,$ to the coordinate of the
observation point $r,$ where this ratio is small -- this makes the
dipole-like approximation of Subsection \ref{dipole}. The simplifying
circumstance that makes this approximation easy to handle is that it so
happens that one needs, as a matter of fact, to solve only the second
Maxwell equation$,$ the one following from the least action principle, while
the first one, $\left[ \mathbf{\nabla \times E}\right] =0,$ is trivially
satisfied. The above general statement concerning the nullification of the
repulsion force at $R=0$ for equal charges is traced at the dynamical level
of Subsection \ref{dipole} \footnote{%
Throughout the paper, Greek indices span Minkowski space-time, Roman indices
span its three-dimensional subspace. Boldfaced letters are three-dimensional
vectors, same letters without boldfacing and index designate their lengths,
except the coordinate vector $\mathbf{x=r}$, whose length is denoted as $r.$
The scalar product is ($\mathbf{r\cdot R)=}x_{i}R_{i},$ the vector product
is $\mathbf{C=}\left[ \mathbf{r\times R}\right] ,$ $C_{i}=\epsilon
_{ijk}x_{i}R_{k}$}

\section{Nonlinear Maxwell equations}

\label{NME}

\subsection{Nonlinear Maxwell equations as they originate from QED}

\label{QED}

It is known that QED is a nonlinear theory due to virtual electron-positron
pair creation by a photon. The nonlinear Maxwell equation of QED for the
electromagnetic field tensor $F_{\nu \mu }\left( x\right) =\partial ^{\mu
}A^{\nu }(x)-\partial ^{\nu }A^{\mu }(x)$ $(\tilde{F}_{\tau \mu }\left(
x\right) $ designates its dual tensor $\tilde{F}^{\mu \nu }=\left(
1/2\right) \varepsilon ^{\mu \nu \rho \sigma }F_{\rho \sigma }$ ) produced
by the classical source \ $J_{\mu }\left( x\right) $ may be written as, see 
\textit{e.g.} \cite{AdoGitSha2016}.%
\begin{equation}
\partial ^{\nu }F_{\nu \mu }\left( x\right) -\partial ^{\tau }\left[ \frac{%
\delta \mathcal{L}\left( \mathfrak{F,G}\right) }{\delta \mathfrak{F}\left(
x\right) }F_{\tau \mu }\left( x\right) +\frac{\delta \mathcal{L}\left( 
\mathfrak{F,G}\right) }{\delta \mathfrak{G}\left( x\right) }\tilde{F}_{\tau
\mu }\left( x\right) \right] =J_{\mu }\left( x\right) \,.  \label{MaxEq}
\end{equation}%
Here $\mathcal{L}\left( \mathfrak{F,G}\right) $ is the effective Lagrangian
(a function of the two field invariants $\mathfrak{F}=\frac{1}{4}F^{\mu \nu
}F_{\mu \nu }$ and $\mathfrak{G}=\left( 1/4\right) \tilde{F}^{\mu \nu
}F_{\mu \nu }),$ of which the generating functional of
one-particle-irreducible vertex functions, called effective action \cite%
{weinberg}, is obtained by the space-time integration as$\ \Gamma \left[ A%
\right] =\int \mathcal{L}\left( x\right) d^{4}x.$ Eq. (\ref{MaxEq}) is the
realization of the least action principle 
\begin{equation}
\frac{\delta S\left[ A\right] }{\delta A^{\mu }\left( x\right) }=\partial
^{\nu }F_{\nu \mu }\left( x\right) +\frac{\delta \Gamma \left[ A\right] }{%
\delta A^{\mu }\left( x\right) }=J_{\mu }\left( x\right) \,,  \label{nm3}
\end{equation}%
where the full action $S\left[ A\right] =S_{\mathrm{Max}}\left[ A\right]
+\Gamma \left[ A\right] $ includes the standard classical, Maxwellian,
electromagnetic action $S_{\mathrm{Max}}\left[ A\right] =-\int \mathfrak{F}%
\left( x\right) d^{4}x$ with its Lagrangian known as $L_{\text{Max}}=-%
\mathfrak{F=}\frac{1}{2}\left( E^{2}-B^{2}\right) $ in terms of the electric
and magnetic fields, $\mathbf{E}$ and $\mathbf{B}.$

Eq. (\ref{MaxEq}) is reliable only as long as its solutions vary but slowly
in the space-time variable $x_{\mu },$ because we do not include the space
and time derivatives of $\mathfrak{F}$ and$\mathfrak{\ G}$ as possible
arguments of the functional $\Gamma \left[ A\right] $ treated approximately
as local$.$ This infrared, or local approximation shows itself as a rather
productive tool \cite{AdoGitSha2016}-- \cite{King}. The calculation of one
electron-positron loop with\ the electron propagator taken as solution to
the Dirac equation in an arbitrary combination of constant electric and
magnetic fields of any magnitude supplies us with a useful example of $%
\Gamma \left[ A\right] $ known as the E-H effective action \cite{Heisenberg}%
. It is valid to the lowest order in the fine-structure constant $\alpha ,$
but with no restriction imposed on the the background field, except that it
has no nonzero space-time derivatives$.$ Two-loop expression of this local
functional is also available \cite{Ritus}.

The dynamical Eq. (\ref{MaxEq}), which makes the "second pair" of Maxwell
equations, may be completed by postulating also their "first pair" 
\begin{equation}
\partial _{\nu }\widetilde{F}^{\nu \mu }\left( x\right) =0\,,
\label{bianchi}
\end{equation}%
whose fulfillment allows using the 4-vector potential $A^{\nu }(x)$ for
representation of the fields: $F_{\nu \mu }\left( x\right) =\partial ^{\mu
}A^{\nu }(x)-\partial ^{\nu }A^{\mu }(x).$ This representation is important
for formulating the least action principle and quantization of the
electromagnetic field. From it Eq. (\ref{bianchi}) follows identically,
unless the potential has singularity like the Dirac string peculiar to
magnetic monopole. In the present paper we keep to Eq. (\ref{bianchi}),
although its denial is not meaningless, as discussed in Ref. \cite%
{AdoGitSha2015}, where a magnetic charge is produced in nonlinear
electrodynamics.

We want now to separate the electrostatic case. This may be possible if the
reference frame exists where all the charges are at rest, $J_{0}\left(
x\right) =J_{0}\left( \mathbf{r}\right) $ . (We denote $\mathbf{r=x).}$Then
in this "rest frame" the spacial component of the current disappears, $%
\mathbf{J}\left( x\right) =0,$ and the purely electric time-independent
configuration $F_{ij}\left( \mathbf{r}\right) =0$ would not contradict to
equation (\ref{MaxEq}). With the magnetic field equal to zero, the invariant 
$\mathfrak{G=}$ $\left( \mathbf{E}\cdot \mathbf{B}\right) $ disappears, too.
In a theory even under the space reflection, to which class QED belongs,
also we have $\left. \frac{\partial \mathcal{L}\left( \mathfrak{F,G}\right) 
}{\partial \mathfrak{G}\left( x\right) }\right\vert _{\mathfrak{G}=0}=0,$
since the Lagrangian should be an even function of the pseudoscalar $%
\mathfrak{G.}$Then we are left with the equation for a static electric field 
$E_{i}=F_{i0}\left( \mathbf{x}\right) $%
\begin{equation}
\partial _{i}F_{i0}\left( \mathbf{r}\right) -\partial _{i}\frac{\delta 
\mathcal{L}\left( \mathfrak{F,}0\right) }{\delta \mathfrak{F}\left( \mathbf{r%
}\right) }F_{i0}(\mathbf{r})=J_{0}\left( \mathbf{r}\right) .\,
\label{static}
\end{equation}

\subsection{\protect\bigskip Model approach}

\label{model}

Equation (\ref{static}) is seen to be the equation of motion stemming
directly from the Lagrangian 
\begin{equation}
L=-\mathfrak{F+}\mathcal{L}\left( \mathfrak{F,}0\right)  \label{L}
\end{equation}%
with the constant external charge $J_{0}\left( \mathbf{r}\right) .$ In the
rest of the paper we shall be basing on this Lagrangian in understanding
that it may originate from QED as described above or, alternatively, be
given \textit{ad hoc} to define a certain model\textit{. }In the latter
case, if treated seriously as applied to short distances near a point charge
where the field cannot be considered as slowly varying, in other words,
beyond the applicability of the infrared approximation of QED outlined
above, the Lagrangian (\ref{L}) may be referred to as defining an extension
of QED to short distances once $\mathcal{L}\left( \mathfrak{F,}0\right) $ is
the E-H Lagrangian (or else its multi-loop specification) restricted to $%
\mathfrak{G}=0$.

It was shown in \cite{CosGitSha2013a} that the important property of
finiteness of the field energy of the point charge is guarantied if $%
\mathcal{L}\left( \mathfrak{F,}0\right) $ in (\ref{L}) is a polynomial of
any power, obtained, for instance, by truncating the Taylor expansion of the
H-E Lagrangian at any integer power of $\mathfrak{F.}$ On the other hand, it
was indicated in \cite{Shishmarev} that a weaker condition is sufficient: if 
$\mathcal{L}\left( \mathfrak{F,}0\right) \mathfrak{\ }$grows with -$%
\mathfrak{F}$ as $\left( -\mathfrak{F}\right) ^{w}$, the field energy is
finite provided that $w>\frac{3}{2}.$ The derivation of this condition is
given in \cite{Ependiev} and in \cite{lecture}. As a matter of fact a more
subtle condition suffices: $\mathcal{L}\left( \mathfrak{F}\right) \sim
\left( -\mathfrak{F}\right) ^{\frac{3}{2}}\ln ^{u}\left( -\mathfrak{F}%
\right) ,$ \ $u>2.$ In what follows any of these sufficient conditions is
meant to be fulfilled.

In the present paper we confine ourselves to the simplest example of the
nonlinearity generated by keeping only quadratic terms in the Taylor
expansion of the E-H Lagrangian in powers of the field invariant $\mathfrak{F%
}$ 
\begin{equation*}
\mathcal{L}\left( \mathfrak{F((}x)\mathfrak{,}0\right) =\frac{1}{2}\left. 
\frac{d^{2}\mathcal{L}\left( \mathfrak{F,}0\right) }{d^{2}\mathfrak{F}}%
\right\vert _{\mathfrak{F}=0}\mathfrak{F}^{2}\mathfrak{(}x),
\end{equation*}%
where the constant and linear terms are not kept, because their inclusion
would contradict the correspondance principle that does not admit changing
the Maxwell Lagrangian $L_{\text{Max}}=-\mathfrak{F}$ for small fields. The
correspondance principle is laid into the calculation of the E-H Lagrangian
via the renormalization procedure.

Finally, we shall be dealing with the model Lagrangian quartic in the field
strength%
\begin{equation}
L=-\mathfrak{F(}x)\mathfrak{+}\frac{1}{2}\gamma \mathfrak{F}^{2}\mathfrak{(}%
x)  \label{quartic}
\end{equation}%
with $\gamma $ being a certain self-coupling coefficient with the
dimensionality of the fourth power of the length, which may be taken as 
\begin{equation*}
\gamma =\left. \frac{d^{2}\mathcal{L}\left( \mathfrak{F,}0\right) }{d^{2}%
\mathfrak{F}}\right\vert _{\mathfrak{F}=0}=\frac{e^{4}}{45\pi ^{2}m^{4}},
\end{equation*}%
where $e$ and $m$ are the charge and mass of the electron, if $\mathcal{L}$
is chosen to be the E-H one-loop Lagrangian. We do not refer to this choice
henceforward. Generalization to general Lagrangians can be also done in a
straightforward way.

The second (\ref{static}) and the first (\ref{bianchi}) Maxwell equations
for the electric field $\mathbf{E}$ with Lagrangian (\ref{quartic}) are

\begin{gather}
\mathbf{\nabla }\cdot \left[ \left( 1+\frac{\gamma }{2}E^{2}(\mathbf{r}%
)\right) \mathbf{E}(\mathbf{r})\right] =j_{0}(\mathbf{r}),  \label{E_div} \\
\mathbf{\nabla }\times \mathbf{E}(\mathbf{r})=0.  \label{rot}
\end{gather}%
Denoting the solution of the linear Maxwell equations as $\mathbf{E}^{lin}(%
\mathbf{r})$ 
\begin{equation}
\nabla \cdot \mathbf{E}^{lin}(\mathbf{r})=j_{0}(\mathbf{r}),\quad \mathbf{%
\nabla }\times \mathbf{E}^{lin}(\mathbf{r})=0,  \label{eqM_lin}
\end{equation}%
we write the solution of (\ref{E_div}), in the following way \cite%
{AdoGitSha2016} -- \cite{CosGitSha2013}%
\begin{equation}
\left( 1+\frac{\gamma }{2}E^{2}(\mathbf{r})\right) \mathbf{E}(\mathbf{r})=%
\mathbf{E}^{lin}(\mathbf{r})+[\mathbf{\nabla }\times \mathbf{\Omega }(%
\mathbf{r})],  \label{E_alg}
\end{equation}%
where the vector function $\mathbf{\Omega }(\mathbf{r})$ may be chosen in
such a way that $\nabla\cdot\mathbf{\Omega }(\mathbf{r})=0.$ Imposing
equation (\ref{rot}) we get 
\begin{equation}
\mathbf{\Omega }(\mathbf{r})=\frac{1}{\nabla ^{2}}[\mathbf{\nabla }\times 
\boldsymbol{\mathcal{E}}(\mathbf{r})]=-\frac{1}{4\pi }\int \frac{[\mathbf{%
\nabla }^{\prime }\times \boldsymbol{\mathcal{E}}(\mathbf{r^{\prime }})]d%
\mathbf{r}^{\prime }}{|\mathbf{r}-\mathbf{r^{\prime }}|},\text{ }
\label{eq_Omega}
\end{equation}%
where we have introduced the auxiliary electric field as the cubic
combination%
\begin{equation*}
\boldsymbol{\mathcal{E}}(\mathbf{r})=\frac{\gamma }{2}E^{2}(\mathbf{r})%
\mathbf{E}(\mathbf{r}).
\end{equation*}%
(In the case of a general Lagrangian that would be a more complicated
function of $\mathbf{E}(\mathbf{r}),$ namely $\boldsymbol{\mathcal{E}}(%
\mathbf{r})=\frac{\delta \mathcal{L}\left( \mathfrak{F,}0\right) }{\delta 
\mathfrak{F}\left( \mathbf{x}\right) }\mathbf{E}(\mathbf{r})).$ From (\ref%
{E_alg}), (\ref{eq_Omega}) it follows that 
\begin{equation}
\mathbf{E}(\mathbf{r})+\boldsymbol{\mathcal{E}}(\mathbf{r})=\mathbf{E}^{lin}(%
\mathbf{r})+[\mathbf{\nabla }\times \mathbf{\Omega }(\mathbf{r})]=\mathbf{E}%
^{lin}(\mathbf{r})+\frac{[\mathbf{\nabla }\times \lbrack \mathbf{\nabla }%
\times \boldsymbol{\mathcal{E}}(\mathbf{r})]]}{\nabla ^{2}},
\end{equation}%
or, in components,%
\begin{equation}
E_{i}(\mathbf{r})=E_{i}^{lin}(\mathbf{r})+\frac{\partial _{i}\partial _{j}}{%
\nabla ^{2}}\frac{\gamma }{2}E^{2}(\mathbf{r})E_{j}(\mathbf{r}).
\label{inteq}
\end{equation}%
In the centre-symmetric case of a single point charge considered in \cite%
{CosGitSha2013a}, \cite{Shishmarev}, the projection operator $\frac{\partial
_{i}\partial _{j}}{\nabla ^{2}}$ in the latter equation is identity ($%
\mathbf{\Omega }(\mathbf{r})=0$), and Eq. (\cite{inteq}) is no longer an
integral equation. The same will be the case in the cylindric-symmetric
problem of two point charges within the approximations to be considered in
the next Section. This simplification makes solution possible. In this case
it is sufficient present the solution of the differential part of Eq. (\ref%
{E_div}) in the form (\ref{E_alg}) setting $\mathbf{\Omega }(\mathbf{r})=0$
in it, then the first Maxwell equation (\ref{rot}) is fulfilled
automatically.

\section{Two-body problem}

\bigskip \label{Appr}

By the two point charge problem we mean the one, where the current $j_{0}(%
\mathbf{r})$ in (\ref{E_div}) is the sum of delta-functions centered in the
positions $\mathbf{r=\pm R}$ of two charges $q_{1}$ and $q_{2}$ separated by
the distance $2R$ (with the origin of coordinates $x_{i}$ placed in the
middle between the charges)%
\begin{equation}
\mathbf{\nabla }\cdot \left[ \left( 1+\frac{\gamma }{2}E^{2}(\mathbf{r}%
)\right) \mathbf{E}(\mathbf{r})\right] =q_{1}\delta ^{3}\left( \mathbf{r-R}%
\right) +q_{2}\delta ^{3}\left( \mathbf{r}+\mathbf{R}\right) .
\label{two-body}
\end{equation}%
In what follows we shall be addressing this equation accompanied by (\ref%
{rot}) for the combined field of two charges.

We shall be separately interested in the force acting between them. The
force $F_{i}=\frac{dP^{0}}{dR_{i}}$ should be defined as the derivative of
the field energy $P^{0}=\int \Theta ^{00}d^{3}x$ stored in the solution of
Eqs. (\ref{two-body}), (\ref{rot}) over the distance between them.

The Noether energy-momentum tensor for the Lagrange density (\ref{L}) is%
\begin{equation}
T^{\rho \nu }=(1-\gamma \mathfrak{F}\left( x\right) )F^{\mu \nu }\partial
^{\rho }A_{\mu }-\eta ^{\rho \nu }L(x).
\end{equation}%
By subtracting the full derivative $\partial _{\mu }\left[ \left( 1-\gamma 
\mathfrak{F}\left( x\right) F^{\mu \nu }\right) A^{\rho }\right] ,$\ equal
to $\left[ \left( 1-\gamma \mathfrak{F}\left( x\right) F^{\mu \nu }\right)
\partial _{\mu }A^{\rho }\right] $ due to the field equations (\ref{MaxEq})
(without the source and with no dependence on $\mathfrak{G}$), the
gauge-invariant and symmetric under the transposition $\rho \leftrightarrows
\nu $ energy-momentum tensor\bigskip \qquad 
\begin{equation}
\Theta ^{\rho \nu }=(1-\gamma \mathfrak{F}\left( x\right) )F^{\mu \nu
}F_{\mu }^{\ \ \rho }-\eta ^{\rho \nu }L(x)  \label{Energy density1}
\end{equation}%
is obtained. This is the expression for the electromagnetic energy proper,
without the interaction energy with the source, the same as in the reference
book \cite{Landau}. When there is electric field alone, the energy density is%
\begin{equation}
\Theta ^{00}=(1+\frac{\gamma E^{2}}{2})E^{2}-\frac{E^{2}}{2}\left( 1+\frac{%
\gamma E^{2}}{4}\right) =\frac{E^{2}}{2}+\frac{3\gamma E^{4}}{8}.
\label{Energy density}
\end{equation}

The integral for the full energy of two charges $P^{0}=\int \Theta
^{00}d^{3}x$ converges since it might diverge only when integrating over
close vicinities of the charges. But in each vicinity the field of the
nearest charge dominates, and we know from the previous publication \cite%
{CosGitSha2013a} (also to be explained below) that the energy of a separate
charge converges in the present model. When the charges are in the same
point, $R=0,$ they make one charge $q_{1}+q_{2},$ whose energy coverges, too.

The energy 
\begin{equation}
P^{0}=\int \Theta ^{00}d^{3}x  \label{P0}
\end{equation}%
is rotation-invariant. Hence it may only depend on the length $R,$ in other
words, be an even function of $\mathbf{R.}$ Then, in the point of
coincidence $\mathbf{R=}0,$ the force $F_{i}=\frac{dP^{0}}{dR_{i}}$ must
either disappear -- if $P^{0}$ is a differentiable function of $R$ \ that
point-- or be infinite -- if not. Crucial to distinguish these cases is the
value of the charge difference $\delta q=q_{2}-q_{1}.$ If the two charges
are equal, $\delta q=0,$ the solution of equation ($\ref{two-body}$) for the
field is an even function of $\mathbf{R,}$ since this equation is invariant
under the reflection $\mathbf{R\rightarrow -R.}$ We shall see in the next
subsection that the linear term in the expansion of the solution in powers
of the small ratio $\frac{\mathbf{R}}{r}$\textbf{\ }is identical zero in
this special case, and so is the linear term of $P^{0}.$

\subsection{\protect\bigskip Large distance case $r\gg R$ (dipole
approximation)}

\bigskip \label{dipole}

We shall look for the solution in the form 
\begin{equation*}
\mathbf{E}=\mathbf{E}^{(0)}+\mathbf{E}^{(1)}+...
\end{equation*}%
where $\mathbf{E}^{(0)}$ and $\mathbf{E}^{(1)}$ are contributions of the
zeroth and first order with respect to the ratio $\frac{\mathbf{R}}{r},$
respectively$.$

The zero-order term is spherical-symmetric, because it corresponds to two
charges in the same point that make one charge, \ 
\begin{equation}
\mathbf{E}^{(0)}=\frac{\mathbf{r}}{r}\mathbf{E}^{(0)}(r).  \label{anz1}
\end{equation}%
Eq. (\ref{rot}) is automatically fulfilled for this form.

Let us write the first-order term $E_{i}^{(1)}$ in the following general
cylindric-symmetric form, linear in the ratio $\frac{\mathbf{R}}{r}$%
\begin{equation}
\mathbf{E}^{(1)}=\mathbf{r}\left( \mathbf{R\cdot r}\right) a(r)+\mathbf{R}%
g(r),  \label{cylform}
\end{equation}%
where $a$ and $g$ \ are functions of the only scalar $r,$ and the cylindric
axis is fixed as the line passing through the two charges. Let us subject (%
\ref{cylform}) to the equation (\ref{rot}) $\nabla \times \mathbf{E}%
^{(1)}=0. $ This results in the relation%
\begin{equation}
a(r)=\frac{1}{r}\frac{\text{d}}{\text{d}r}g(r),  \label{connection}
\end{equation}%
provided that the vectors $\mathbf{r,R}$ are not parallel. We shall see that
with the ansatzes (\ref{cylform}) and (\ref{anz1}) equation (\ref{E_alg})
can be satisfied with the choice $\mathbf{\Omega }(\mathbf{r})=0:$ 
\begin{equation}
\left( 1+\frac{\gamma }{2}E^{2}(\mathbf{r})\right) \mathbf{E}(\mathbf{r})=%
\mathbf{E}^{lin}(\mathbf{r}),  \label{without}
\end{equation}%
namely, we shall find the coefficient functions $a,$ $g$ from Eq. (\ref%
{without}) and then ascertain that the relation ( \ref{connection}) is
obeyed by the solution.

\bigskip The inhomogeneity in (\ref{without})%
\begin{equation*}
\mathbf{E}^{lin}\mathbf{(r})=\frac{q_{1}}{4\pi }\frac{\mathbf{r-R}}{|\mathbf{%
r-R}|^{3}}+\frac{q_{2}}{4\pi }\frac{\mathbf{r+R}}{|\mathbf{r+R}|^{3}}
\end{equation*}%
satisfies the linear ($\gamma =0)$ limit of equation \ (\ref{two-body}) 
\begin{equation}
\mathbf{\nabla }\cdot \mathbf{E}^{lin}(\mathbf{r})=q_{1}\delta ^{3}\left( 
\mathbf{r-R}\right) +q_{2}\delta ^{3}\left( \mathbf{r}+\mathbf{R}\right)
\end{equation}%
and also (\ref{rot}). The inhomogeneity is expanded in $\frac{\mathbf{R}}{r}$%
as 
\begin{eqnarray}
\mathbf{E}^{lin}\mathbf{(r}) &=&\frac{(q_{1}+q_{2})}{4\pi r^{2}}\frac{%
\mathbf{r}}{r}+\frac{\left( q_{2}-q_{1}\right) }{4\pi r^{2}}\left( \frac{%
\mathbf{R}}{r}-3\frac{\mathbf{r}}{r}\frac{\left( \mathbf{R\cdot r}\right) }{%
r^{2}}\right) +...=  \notag \\
&=&\frac{(q_{1}+q_{2})}{4\pi r^{2}}\frac{\mathbf{r}}{r}+\frac{1}{4\pi }%
\left( \frac{\mathbf{d}}{r^{3}}-\frac{3\left( \mathbf{d\cdot r}\right) }{%
r^{5}}\mathbf{r}\right) \mathbf{+...,}  \label{inhom}
\end{eqnarray}%
where $\mathbf{d=}\left( q_{2}-q_{1}\right) \mathbf{R}$ is the dipole
moment, while the dots stand for the disregarded quadrupole and higher
multipole contributions.

The zero-order term satisfies the equation%
\begin{equation}
\left( 1+\frac{\gamma }{2}E^{(0)2}(r)\right) E^{(0)}(r)=\frac{(q_{1}+q_{2})}{%
4\pi r^{2}},  \label{zeroord}
\end{equation}%
with the first term of expansion (\ref{inhom}) taken for inhomogeneity. This
is an algebraic (not differential) equation, cubic in the present model (\ref%
{quartic}), solved explicitly for the field $E^{(0)}$ as a function of $r$
in this case, but readily solved for the inverse function $r(E^{(0)})$ in
any model$,$ which is sufficient for many purposes. Even without solving it
we see that for small $r\ll \gamma ^{\frac{1}{4}}$ the second term in the
bracket dominates over the unity, therefore the asymptotic behavior in this
region follows from (\ref{zeroord}) to be%
\begin{equation*}
E^{(0)}(r)\sim \left( \frac{q_{1}+q_{2}}{2\pi \gamma }\right) ^{\frac{1}{3}%
}r^{-\frac{2}{3}}.
\end{equation*}%
This weakened -- as compared to the Coulomb field $\frac{q_{1}+q_{2}}{4\pi }%
r^{-2}$ -- singularity is not an obstacle for convergence of the both
integrals in (\ref{P0}), (\ref{Energy density}) for the proper field energy
of the equivalent point charge $q_{1}+q_{2}.$

With the zero-order equation (\ref{zeroord}) fulfilled, we write \ a linear
equation for the first-order correction $\mathbf{E}^{(1)}$ from (\ref%
{without}), to which the second, dipole part in (\ref{inhom}) serves as an
inhomogeneity 
\begin{equation*}
\mathbf{E}^{(1)}=\frac{\left( q_{2}-q_{1}\right) }{r^{2}}\left( \frac{%
\mathbf{R}}{r}-3\frac{\mathbf{r}}{r}\frac{\left( \mathbf{R\cdot r}\right) }{%
r^{2}}\right) -\frac{\gamma }{2}\left[ 2\left( \mathbf{E}^{(1)}\cdot \mathbf{%
E}^{(0)}\right) \mathbf{E}^{(0)}+E^{(0)2}\mathbf{E}^{(1)}\right] .
\end{equation*}%
This equation is linear and it does not contain derivatives. We use (\ref%
{cylform}) as the ansatz. After calculating%
\begin{equation*}
2\left( \mathbf{E}^{(1)}\cdot \mathbf{E}^{(0)}\right) \mathbf{E}%
^{(0)}+E^{(0)2}\mathbf{E}^{(1)}=\mathbf{r}E^{(0)2}\frac{\left( \mathbf{%
R\cdot r}\right) }{r^{2}}\left( 2g+3r^{2}a\right) +\mathbf{R}gE^{(0)2},
\end{equation*}%
we obtain two equations, along $\mathbf{R}$ and $\mathbf{r,}$ with the
solutions ($\delta q$=$q_{2}-q_{1},$ $Q=$ $q_{2}+q_{1}):$ 
\begin{equation}
g=\frac{\delta q}{r^{3}}\frac{1}{1+\frac{\gamma }{2}E^{(0)2}}=\frac{\delta q%
}{Qr}E^{(0)},  \label{gg}
\end{equation}

\begin{equation}
a=-\frac{\delta q}{r^{5}}\frac{3+\frac{5\gamma }{2}E^{(0)2}}{\left( 1+\frac{%
\gamma }{2}E^{(0)2}\right) \left( 1+\frac{3\gamma }{2}E^{(0)2}\right) }
\label{a}
\end{equation}
From (\ref{zeroord}) we obtain 
\begin{equation*}
\frac{\text{d}}{\text{d}r}E^{(0)}=-\frac{2Q}{r^{3}\left( 1+\frac{\gamma }{2}%
E^{(0)2}\right) }-\frac{\gamma E^{(0)2}}{1+\frac{\gamma }{2}E^{(0)2}}\frac{%
\text{d}}{\text{d}r}E^{(0)}.
\end{equation*}%
Hence 
\begin{equation}
\frac{\text{d}}{\text{d}r}E^{(0)}=-\frac{2Q}{r^{3}\left( 1+\frac{3\gamma }{2}%
E^{(0)2}\right) }.  \label{dE/dr}
\end{equation}%
With the help of this relation the derivative of (\ref{gg}) can be
calculated to coincide with $\left( \ref{a}\right) $ times $r.$ This proves
Eq. (\ref{connection}) necessary to satisfy the first Maxwell equation (\ref%
{rot}).

By comparing this with (\ref{a}) we see that Eq. (\ref{connection})
necessary to satisfy the first Maxwell equation (\ref{rot}) has been proved.

Finally, relations (\ref{gg}) and (\ref{a}) as substituted in the general
cylindric covariant decomposition (\ref{cylform}) give the linear in $\frac{%
\mathbf{R}}{r}$(\ref{a}) correction $\mathbf{E}^{(1)}$ to the zero-order
field $\mathbf{E}^{(0)}$, subject to the equation (\ref{zeroord}), in terms
of $E^{(0)},$which is explicitly known in our special model. These results
may be considered as giving nonlinear correction to the electric dipole
field (the second term in (\ref{inhom})) due to nonlinearity.

Coming back to the discussion on the repulsion force we have to analyze the
contribution of the found linear term $\mathbf{E}^{(1)}$ into the energy.
The contribution of $\mathbf{E}^{(1)}$ into the energy density (\ref{Energy
density}) linear in $\mathbf{R}$ contains the factor $\left( \mathbf{E}%
^{(1)}\cdot \mathbf{E}^{(0)}\right) =\left( \mathbf{E}^{(1)}\cdot \frac{%
\mathbf{r}}{r}\right) E^{(0)}.$ According to the result (\ref{cylform}) this
factor is linear with respect to the scalar product $\left( \mathbf{R\cdot r}%
\right) =Rr\cos \theta .$ It would give zero contribution into the energy (%
\ref{P0}) due to the angle integration. This does not imply, however, that
the force at the point of coincidence $\mathbf{R=}0$ is zero, because the
linear contribution into the integrand in (\ref{P0}) would create divergence
of the integral (\ref{P0}) near $r=0.$ The interchange of the integration
over $r$ and of the limiting transition $\frac{R}{r}\rightarrow 0$ is not
permitted. In the region $r<R$ of integration the linear approximation in
the ratio $\frac{R}{r}$ is not relevant. This region gives the infinite
contribution into the repulsion force between two charges when the approach
each other infinitely close. The case where these charges are equal, $%
q_{1}-q_{2}=0,$ is different. Then the solution for $\mathbf{E}^{(1)}$ is
just zero, and we confirm the conclusion made above following general
argumentation that equal point charges do not repulse when their positions
coincide.

\section*{Acknowledgements}

Supported by RFBR under Project 14-02-01171, and by the TSU Competitiveness
Improvement Program, by a grant from \textquotedblleft The Tomsk State
University D. I. Mendeleev Foundation Program\textquotedblright .


\end{document}